\documentstyle[12pt,psfig]{article}
\def\reference{\parskip 0pt\par\noindent\hangindent 0.5 truecm}

\def\ltsima{$\; \buildrel < \over \sim \;$}
\def\simlt{\lower.5ex\hbox{\ltsima}} 
\def\gtsima{$\; \buildrel > \over \sim \;$}
\def\simgt{\lower.5ex\hbox{\gtsima}} 

\begin{document}
%
%
\title{X-ray Variability of Blazars}
%


\author{Elena Pian}

\date{}
\maketitle

{\center
INAF, Astronomical Observatory of Trieste, Via G.B. Tiepolo 11, I-34131
Trieste, Italy\\pian@ts.astro.it\\[3mm]
}

%
\begin{abstract}

Critical progress in our understanding of high energy emission
from AGN has been determined in the last 10 years by X-ray
monitoring campaigns with many space missions, notably ROSAT,
ASCA, RXTE, BeppoSAX, and XMM, often in conjunction with
observations at other frequencies. The emphasis of the present
review is on recent findings about X-ray variability of blazars.
Among AGN, these exhibit the largest amplitude variations of the
X-ray emission, often well correlated with variations at higher
energies (GeV and TeV radiation).  The accurate sampling of the
X-ray spectra over more than three decades in energy, made
possible by the wide energy range of BeppoSAX, has also shown
strong spectral variability in blazar active states, suggesting
extreme electron energies and leading to the identification of a
class of `extreme synchrotron' sources.

\end{abstract}

{\bf Keywords:} radiation mechanisms: non-thermal --- X-rays: galaxies


\bigskip

%
%

\section{Introduction}

The multiwavelength continuum emission of blazar type AGN is
dominated by non-thermal radiation in a relativistic jet pointing
close to the line of sight. Magnification of the light intensity
and variation amplitude, and foreshortening of intrinsic
variability time scales, due to aberration effects, make blazars
the most luminous and variable AGN at all frequencies and the most
direct probe of galactic nuclear continuum emission. The amplitude
of flux density variations increases as a function of frequency.
Sensitive X-ray observations of bright blazars are therefore
effective in measuring temporal and spectral properties on time
scales as short as hours or even minutes. X-ray variability on the
shortest time scales allows us to derive the sizes of the inner
nuclear regions and to trace the kinematics of the jet, and
thereby study how the relativistic particles are accelerated and
injected in the emission region and how efficiently they cool
thereafter and escape. Combined with simultaneous observations at
other wavelengths, X-ray light curves carry information on
multiwavelength spectral variability and on temporal delays of
emission at different frequencies. These can be used to test the
models for the production of the blazar continuum.

\begin{figure}
\begin{center}
\begin{tabular}{cc}
\psfig{file=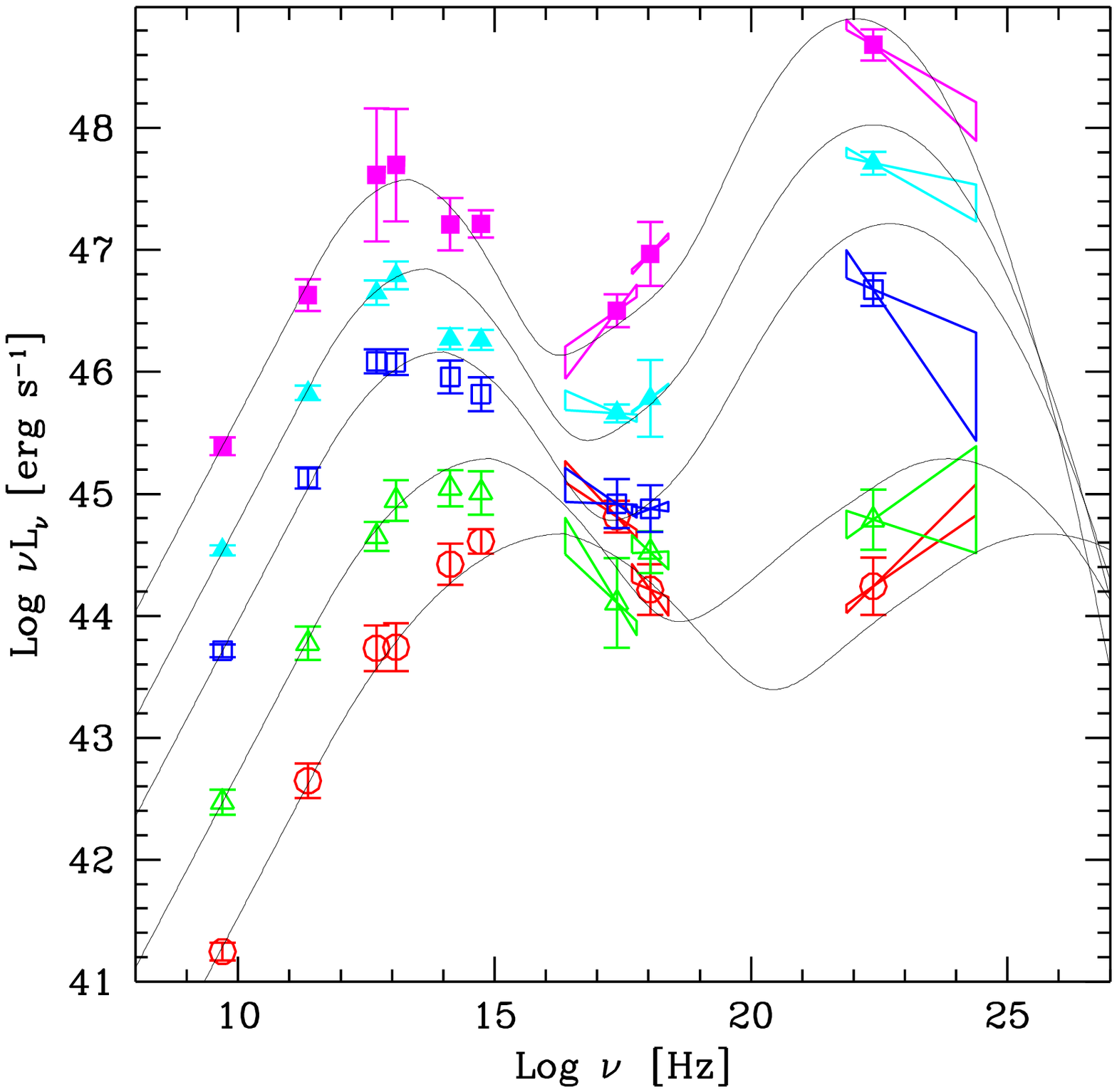,height=6.5cm,clip=}
\psfig{file=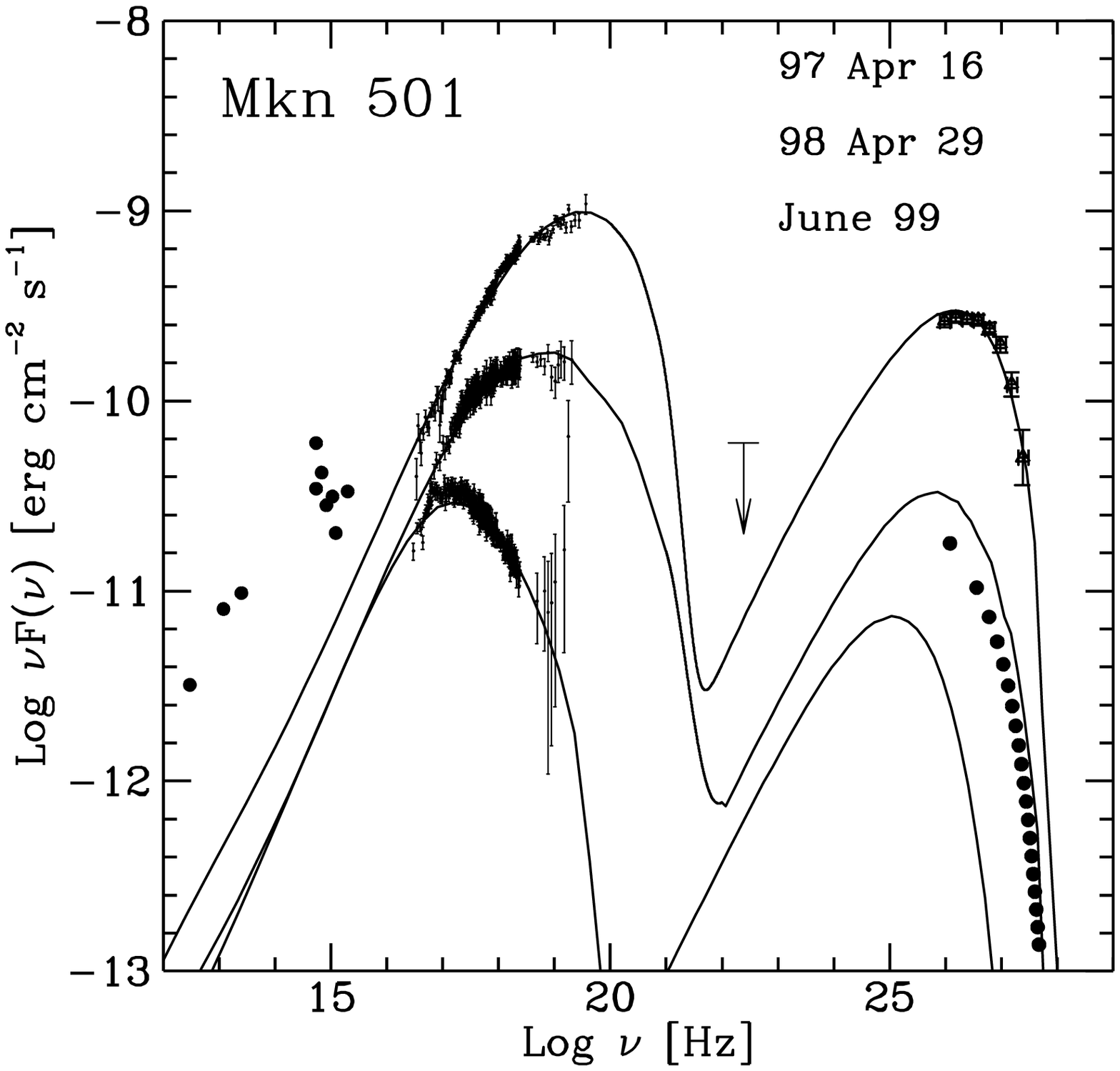,height=7cm,clip=}
\end{tabular}

\caption{Left: Average SEDs of blazars binned according to radio
luminosity (Fossati et al.\ 1998). The curves represent the
parameterisation proposed by Donato et al.\ (2001). Right: SEDs of
Mkn~501. The solid curves are spectra calculated with the
homogeneous SSC model (see Tavecchio et al.\ 2001 for details on
the data and on the model parameters). The filled circles at TeV
energies represent the HEGRA 1998--1999 data (Aharonian et al.\
2001). The assumed emitting region is homogeneous and the general
aim of the modelling is to reproduce the observed variability with
a minimum number of physical parameter changes.  These
approximations may not be totally rigorous and valid (i.e.\ the
emitting region could have a frequency dependent size, with lower
frequency radiation being produced in larger regions). This may
account for the departure of the model from data at optical and
longer wavelengths.  Moreover, these may be contaminated by the
host galaxy emission (from Tavecchio et al.\ 2001).}

\end{center}
\end{figure}

The spectral energy distributions (SEDs) of blazars  are typically
double-peaked (in a $\nu f_\nu$ representation), with a first
component attributed to synchrotron radiation, the maximum of
which is located at IR to soft X-ray frequencies, and a second
component correspondingly peaking at GeV to TeV energies, produced
through inverse Compton scattering (Figure 1, left). According to
whether the peak of their first component is at IR or UV/soft
X-ray wavelengths, the sources have been defined as Low-Energy or
High-Energy Peaked Blazars (Padovani \& Giommi 1995), or as `red'
or `blue' blazars (Fossati et al.\ 1998).  However, rather than
exhibiting a neat distinction between red and blue sources, the
blazar population presents a continuity of properties between
these two extrema, all correlating with the bolometric luminosity.

High energy peaked blazars have lower bolometric luminosities and
smaller ratios between high energy and low energy components
(i.e.\ smaller inverse Compton dominance). Ghisellini et al.\
(1998) identify the ambient (broad emission line region or
accretion disk) photon density as the parameter governing these
correlations: in sources with lower ambient photon densities (blue
blazars) the high energy electrons cool via scattering off
synchrotron photons (synchrotron self-Compton mechanism, SSC),
while for high ambient photon densities the inverse Compton
scattering off external radiation prevails (red blazars).

In blue blazars the radiation efficiency is smaller than in red
ones, and therefore in the former sources electrons can attain
higher energies, thus synchrotron spectra peak at higher
frequencies, and the overall luminosity is lower than in red
blazars (however, Urry 1999 discusses some caveats against this
scenario; see also Giommi, Menna, \& Padovani 1999a).  
The difference in synchrotron luminosities
may be
determined by different degrees of relativistic beaming. Boettcher
\& Dermer (2002) interpret the increasing spectral hardness with
decreasing luminosity of blazars in an evolutionary scenario of
increasing reduction of the black hole accretion power with time
(see also Maraschi 2001).

In the X-ray spectral range the two contributions from synchrotron
and inverse Compton components overlap, with the former dominating
in blue blazars, and the latter in red blazars (see Donato et al.\
2001 for a recent comprehensive review of blazar X-ray spectra).
Therefore, in blue blazars photons up to 10 keV, and occasionally
100 keV or more, especially in outburst, represent the highest
energy tail of the synchrotron radiation.  Being produced by the
most energetic electrons, they are most sensitive to variations in
the acceleration and cooling processes, and are thus expected to
exhibit the most rapid and largest amplitude variability.  In a
few of these blazars the inverse Compton component, produced
through SSC, is detected up to  TeV energies.

In red blazars, the X-ray spectrum is still due to SSC, while external Compton
dominates at the higher frequencies (Takahashi, Madejski, \& Kubo 1999). In the last
10 years, the blazar monitoring by X-ray satellites ROSAT, ASCA, RXTE, BeppoSAX, and
recently XMM, has produced many results on variability, as reviewed, e.g.\ by
Ulrich, Maraschi, \& Urry (1997), Urry (1999), Sambruna (1999), McHardy (1999),
Maraschi et al.\ (2000), and Maraschi \& Tavecchio (2001).  The best studied red
blazars present X-ray variability correlated with optical and gamma rays, although
the variation amplitude is usually smaller than in gamma rays (e.g., Wehrle et al.
1998; Ghisellini et al. 1999). The hard X-ray spectral variability is modest
(Sambruna et al. 1997; Lawson, McHardy, \& Marscher 1999; Hartman et al. 2001;
Ballo et al. 2002).
For some objects (intermediate blazars) the X-ray observations have
simultaneously sampled the synchrotron and the SSC components at soft and hard
X-rays, respectively (Giommi et al.\ 1999b; Tagliaferri et al.\ 2000; Ravasio et
al.\
2002; Tanihata et al.\ 2000).

Blue blazars present a higher amplitude X-ray emission variability than
red blazars (factors of up to 30 with respect to quiescence),
accompanied by different spectral variability (compare the
remarkable changes of spectral slope and synchrotron peak
frequency in Mkn~501, Pian et al.\ 1998, with the modest peak
frequency variations in PKS~2005--489, Perlman et al.\ 1999;
Tagliaferri et al.\ 2001). In the present paper I will summarise
some of the most recent results obtained for the brightest and
best monitored BL Lacs Mkn~501, Mkn~421, and PKS~2155--304, all of
which belong to the class of blue blazars.

\section{Mkn 501}

This source has been observed by BeppoSAX in April 1997,
April--May 1998, and June 1999.  On the first occasion it
exhibited an outburst accompanied by dramatic spectral hardening:
the synchrotron peak energy was observed to reach 100 keV or more,
i.e.\  almost three orders of magnitude larger than in its
quiescent state (Catanese et al.\ 1997; Pian et al.\ 1998). The
following year, the peak energy was found at 20 keV and the
emission intensity above 1 keV was lower.  The fact that, despite
the radiation losses, the synchrotron peak was still at such high
energies one year after the outburst indicates the presence of
efficient, continuously active mechanisms of particle acceleration
in this source.

In June 1999 BeppoSAX recorded an `inactive' state, i.e.\ similar
to the historical one, both in emission level and synchrotron peak
energy ($\sim$0.5 keV), and characterised by modest (amplitude of
10-20\%) intraday flux variability (Tavecchio et al.\ 2001).  In
Figure 1, right, the X-ray spectra are shown together with the
simultaneous TeV observations.  The X-ray spectra are best
described by curved and continuously steepening laws.  Their
variability can be reproduced by changes of the electron break
energy only (i.e.\ the energy of the electrons which emit at the
spectral maximum). 
As frequently noted in blazars (see also Section 3), the X-ray spectral
index and peak
energy correlate well with the source intensity, flatter index and higher
peak energy
corresponding to brighter states.

The good correlation between intensity and spectral index holds
also when previous GINGA and ASCA observations are considered
together with BeppoSAX data (Kataoka et al.\ 1999). The X-ray and
TeV intensities correlate well with no measurable time lag larger
than one day. This suggests that the same electron population is
responsible for the X- and gamma-ray emission through the SSC
mechanism in a homogeneous region (Tavecchio, Maraschi, \&
Ghisellini 1998; Tavecchio et al.\ 2001; Petry et al.\ 2000). The
ensuing prediction for the correlated behaviour of X-ray and TeV
flux densities in different states (Tavecchio et al.\ 2001) is in
good agreement with the simultaneous RXTE and TeV HEGRA
observations in 1997 (see Figure 6 in Krawczynski et al.\ 2000).

Strong correlation between spectral hardness and intensity has
been observed by RXTE as well in June 1998, when a rapid shift in
synchrotron peak energy was also detected during a brightening.
The simultaneous X-ray and TeV flux and spectral variability
points, again, to the SSC origin of the multiwavelength emission
(Sambruna et al.\ 2000).

In July 1997, when Mkn~501 was close to its historically largest
X-ray intensity, RXTE observations of the source revealed unusual
anti-correlation between spectral hardness and intensity (i.e.\
flatter spectrum for dimmer intensity). At that epoch the source
exhibited variations of up to 30\% over a few hours (Lamer \&
Wagner 1998).  Remarkably rapid variability has been detected
during the May 1998 RXTE campaign, when the blazar flared by
factors 1.6 and 2 in $\sim$200 seconds in the 2--10 keV range and
10--15 keV, respectively (Catanese \& Sambruna 2000). With the
event is associated significant spectral variability.

\section{Mkn 421}

A 1 day flare was detected in this source simultaneously in X-rays
by BeppoSAX and at TeV energies by the Whipple Cerenkov telescope
in April 1998 at the beginning of a 10 day multiwavelength
monitoring campaign, involving also the ASCA, EUVE, and RXTE
satellites (Takahashi et al.\ 2000). The X-ray and TeV light
curves are well correlated with no temporal lag larger than 1.5
hours (Figure 2, left), indicating that the radiation at the two
frequencies is produced by the same electron population via SSC
(Maraschi et al.\ 1999). The observed difference in the flare
decay times in X-rays and TeV may be due to light travel time
effects adding to radiative time scales (Chiaberge \& Ghisellini
1999), or to the inhomogeneity of the emitting region (see also
the simultaneous RXTE and TeV observations of June 2000 by
Krawczynski et al.\ 2001).

\begin{figure}
\begin{center}
\begin{tabular}{cc}
\psfig{file=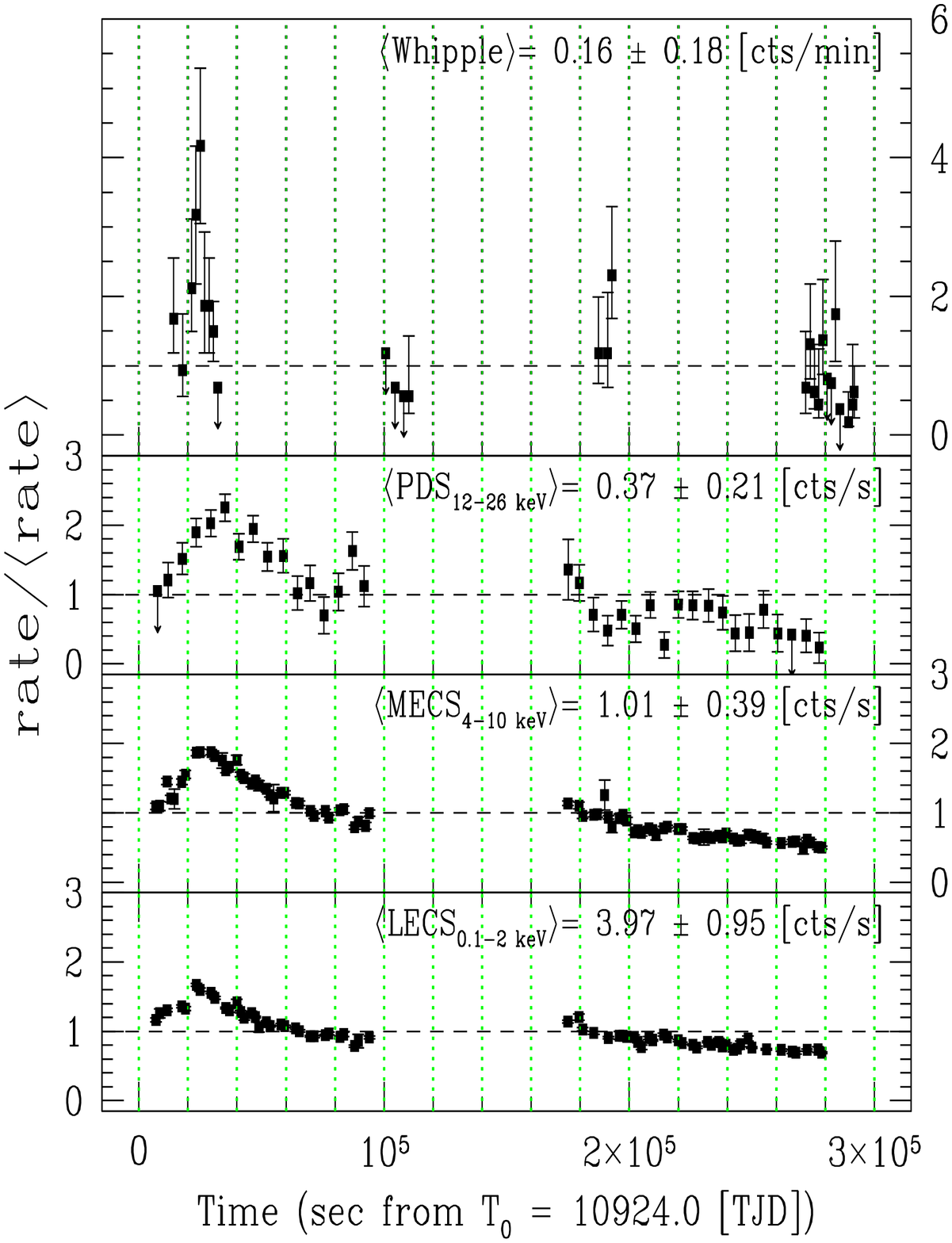,height=7cm,clip=}
\psfig{file=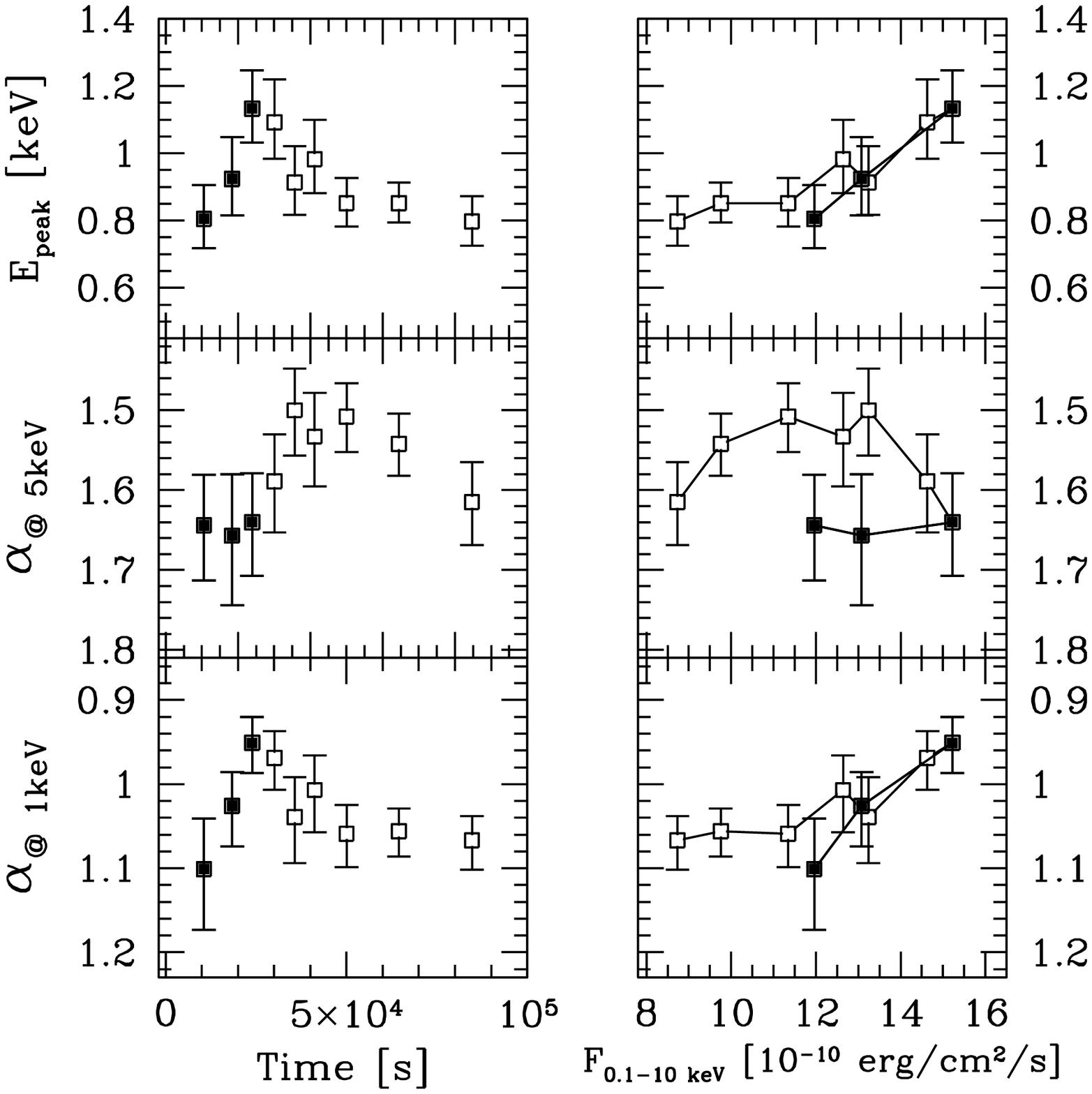,height=7cm,clip=}
\end{tabular}

\caption{April 1998 flare of Mkn~421. Left: light curves at X-ray
(BeppoSAX) and TeV (Whipple) energies on 21--24 April 1998.  All
count rates are normalised to their respective averages (from
Maraschi et al.\ 1999). Right, top to bottom: the energy of the
peak of the synchrotron component, and the photon spectral indices
at 5 and at 1~keV versus time (left column) and de-absorbed
0.1--10~keV intensity (right column).  The solid black symbols
represent the behaviour before the top of the flare, while empty
dots represent the spectral evolution during the decay. The data
points are connected to show the time sequence. A shift of the
synchrotron peak and a delay in the response of the spectrum at
5~keV can be clearly seen (from Fossati et al.\ 2000b).}

\end{center}
\end{figure}

Within the X-ray band the flare decay is achromatic, if a
stationary underlying emission component is postulated (Fossati et
al.\ 2000a). The temporally resolved spectra during the X-ray
flare allowed Fossati et al.\ (2000b) to fully map, for the first
time in any blazar, the increase of the synchrotron peak energy
during the rising part of the flare and its subsequent decrease
during the decay (Figure 2, right). Unlike in Mkn~501, where the
synchrotron peak during the outburst can reach 100 keV or more, in
Mkn~421 the maximum energy reached by the synchrotron peak is 1
keV.  An indication that it may have shifted up to 10 keV or
beyond is present in May 2000 BeppoSAX data, when the source was
also detected in its historical maximum X-ray brightness ($F_{2-10
keV} = 1.2 \times 10^{-9}$ erg s$^{-1}$ cm$^{-2}$, Fossati 2001).
It is interesting to note that Mkn~421 presents larger amplitude
short term variability than Mkn~501, which suggests that maximum
synchrotron peak energy and short time scale variability may have
complementary roles in dissipating the jet energy.

It is generally observed that the X-ray intensity and spectral
index variations in blazars are correlated, in the sense of
spectral hardening during brightening.  It has been noted for some
time (see e.g.\ Sembay et al.\ 1993) that the changes describe
different paths in the index--intensity plane, according to
whether they refer to the increasing or decreasing portion of a
flare. This is seen also during the April 1998 flare of Mkn~421,
which exhibits an `anti-clockwise' pattern of intensity vs
spectral index (middle and bottom panels of the right side of
Figure 2, right). For Mkn~421 this is less common than the
opposite `clockwise' pattern (Takahashi et al.\ 1996; Malizia et
al.\ 2000).

The emission of soft X-rays ($\simlt$2 keV) is generally well
correlated with and lags that of hard X-rays ($\simgt$4 keV) by a
variable interval (from 3--4 ks down to $\sim$300~s) as determined
with cross-correlation methods (Malizia et al.\ 2000; Takahashi et
al.\ 2000; Brinkmann et al.\ 2001). The soft lag is interpreted as
a consequence of the synchrotron cooling, which is more rapid at
higher energies.  The reversal of this temporal correlation has
been observed in 1997 and April 1998 (Treves et al.\ 1999; Fossati
et al.\ 2000a), suggesting that hard lags may be associated with
anti-clockwise intensity spectrum patterns, while soft lags are
associated with clockwise patterns. Hard lags may originate when
the acceleration time scale approaches the cooling time scale, so
that electrons appear first at low energies and gradually build up
at higher energies (Kirk, Rieger, \& Mastichiadis 1998; Takahashi
et al.\ 2000).

We note however that Edelson et al.\ (2001) have put into question
temporal lags of the order of $\sim$1 hour, when obtained by
applying cross-correlation methods to light curves measured by
satellites with low Earth orbits (like ASCA and BeppoSAX).

Mkn~421 flares daily, with a characteristic time scale of 12
hours, as determined from the structure function analysis of the
ASCA data in April 1998 (Takahashi et al.\ 2000; a similar result
has been obtained with previous ASCA observations in 1994,
Takahashi et al.\ 1996, and with XMM observations in May 2000,
Brinkmann et al.\ 2001).  However, 50\% variations of 4--5 hours
time scale are typical (Guainazzi et al.\ 1999; Fossati et al.\
2000a; Krawczynski et al.\ 2001). The symmetric shape of the X-ray
flares detected by BeppoSAX and ASCA imply that the cooling time
scales are shorter than the light crossing time through the
emitting region, so that the latter dominates the variability.

\section{PKS 2155--304}

A blue, relatively nearby ($z \simeq 0.1$) blazar, and one of the
brightest at UV and X-ray frequencies, PKS~2155--304 was expected
to be a TeV emitter and has indeed been detected at these energies
in September--November 1996 by the Cerenkov telescope Mark~6
(Narrabri, Australia) and, with a larger intensity, simultaneously
with EGRET and BeppoSAX in November 1997 (Chadwick et al.\ 1999;
Chiappetti et al.\ 1999; Vestrand \& Sreekumar 1999). The X-ray
state observed by BeppoSAX is one of the highest ever detected,
though lower than that observed by RXTE before the start of the
BeppoSAX campaign. A pure SSC model in a homogeneous region
accounts well for the X-ray and higher energy radiation, with
variability produced solely by changes in electron break energy
(Figure 3).

\begin{figure}
\begin{center}
\begin{tabular}{cc}
\psfig{file=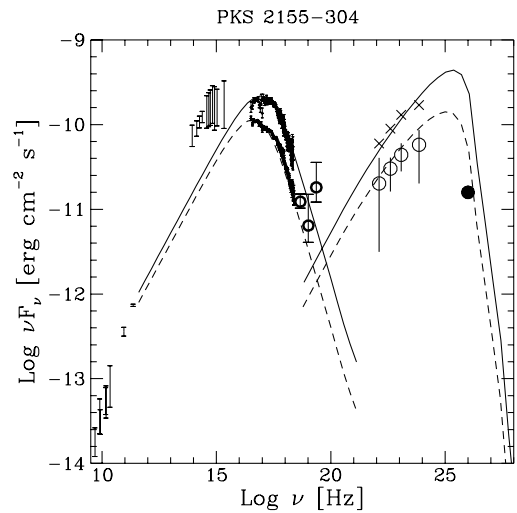,height=8cm,clip=}
\end{tabular}

\caption{ SEDs of PKS~2155--304. The BeppoSAX spectra in the high
and low states and the TeV point refer to simultaneous
observations in November 1997. The vertical bars encompass the
range between the minimum and maximum value in a compilation of
radio, optical, and UV data.  The X-ray spectra are fitted with a
homogeneous SSC model (solid and dashed curves, respectively).  A
comment similar to that made for Mkn~501 (Figure 1, right) applies
here to the inconsistency of the model with UV/optical data (from
Chiappetti et al.\ 1999).}

\end{center}
\end{figure}

A comparative temporal analysis on the X-ray light curves of
PKS~2155--304 in 1994 (ASCA, Kataoka et al.\ 2000), 1996, and 1997
(BeppoSAX, Giommi et al.\ 1998; Chiappetti et al.\ 1999) has been
accomplished by Zhang et al.\ (1999), who found power density
spectra of slope similar to those of Seyfert galaxies ($f^{-1.7}$,
e.g.\ Edelson \& Nandra 1999). The minimum variability time scale
determined by the power density spectrum analysis in the 1.5--10
keV range is $\sim$1000 s, consistent with the minimum measured
doubling time of the flux.

The degree of variability is found to be a function of brightness,
with shorter doubling times and larger fractional variability
parameter for brighter states.  Similarly, the temporal delays
between soft and hard X-rays, determined with different
correlation methods, are shorter for higher source intensity (see
however Section 3 and Edelson et al.\ 2001).

Both positive and negative temporal lags between soft and hard
X-rays have been measured by RXTE for PKS~2155--304 during
successive flares in May 1996, and correspondingly a clockwise or
anti-clockwise loop, respectively, is seen in the intensity vs
hardness ratio plane (Sambruna 1999). This, as in the case of
Mkn~421, can be interpreted in terms of a different duration of
the acceleration process with respect to the electron cooling
(Kirk et al.\ 1998).

The variable temporal lags between soft and hard X-ray emission
may be related to a different geometry of the emitting region in
different states of the source (Zhang et al.\ 1999), requiring a
time-dependent SSC model to reproduce the SED (Kataoka et al.\
2000). Alternatively, Georganopoulos \& Marscher (1998) have
proposed that the different temporal lags observed in 1991 and
1994 between light curves of PKS~2155--304 at UV and X-ray
frequencies (Edelson et al.\ 1995; Urry et al.\ 1997) can be
reproduced by flaring components superimposed on a similar steady
state of the jet, and produced by different energy distributions
of freshly injected electrons.

It is striking that in this source, similar patterns of X-ray
variability are seen, spaced apart by many years (Brinkmann et
al.\ 2000), suggesting recurrent conditions in the mechanism of
jet powering and activity development. This also weakens the
microlensing hypothesis as the origin of intraday variability in
this source (Urry et al.\ 1993).

\section{Conclusions}

The X-ray campaigns, coordinated with TeV telescopes, have
determined a crucial progress toward our understanding of blazars
and have shown the importance of X-ray instruments with wide
spectral range and good temporal resolution. The correlated
variability at X-ray and TeV energies can be reproduced by the
changes of a single parameter, namely the break energy of the
relativistic electron distribution.  In some sources, especially
(but not necessarily) during outburst, this parameter can attain
extremely high values ($\gamma_b > 10^6$), resulting in
synchrotron peaks beyond 1 keV (Mkn~421), and even up to 50--100
keV, as in Mkn~501, 1ES 2344+514 (Giommi, Padovani, \& Perlman
2000), and 1ES 1426+428 (Costamante et al.\ 2001).

More detections of these `extreme synchrotron blazars' are
expected with the advent of the INTEGRAL satellite for soft
gamma-ray astronomy.  These sources are expected to be TeV
emitters, and therefore should be suitable targets for the next
generation of sensitive Cerenkov telescopes.  Simultaneous
observations at X-ray and TeV energies should allow us to measure
the correlation of the light curves on shorter time scales than it
has been possible so far and identify the relation between source
intensity  and variable temporal lags. Similar coordinated
observations of red blazars, which are not expected to be strong
TeV emitters, will receive renewed interest after the launch in
the near future of satellites sensitive in the MeV--GeV range,
i.e.\ AGILE and GLAST.

Finally, longer and denser monitoring of blazars will allow a
better test of the hadronic models of blazar emission. These
predict gamma rays to be due to proton synchrotron radiation
(Aharonian 2000), or to pion production by accelerated protons
interacting with turbulent fluctuations in the magnetic field
(Mannheim 1998), or with ambient radiation fields (Donea \&
Protheroe 2002).  In hadronic models, correlated variability at
X-ray and TeV energies may arise from less restrictive sets of
parameters than in leptonic models (SSC, or synchrotron and
external Compton). However, hadronic and leptonic models are
clearly separated in the parameter space, so that it should be
possible to distinguish them via a model-independent estimate of
the magnetic field and plasma Doppler factor (Rachen 2000).

%
%




\section*{Acknowledgments}

I thank A.\ Celotti, L.\ Chiappetti, G.\ Fossati, G.\ Ghisellini,
F.\ Tavecchio, and A.\ Treves for valuable input, L.\ Maraschi for
a careful reading of the manuscript, and the workshop organisers
for a pleasant, stimulating, and constructive meeting.

\section*{References}






\reference Aharonian, F.\ 2000, New Astronomy, 5, 377

\reference Aharonian, F., et al.\ 2001, ApJ, 546, 898

\reference Ballo, L., et al. 2002, ApJ, 567, 50

\reference Boettcher, M., \& Dermer, C. D. 2002, ApJ, 564, 86 

\reference Brinkmann, W., Gliozzi, M., Urry, C.M., Maraschi, L., \&
Sambruna, R. 2000, A\&A, 362, 105               

\reference Brinkmann, W., et al.\ 2001, A\&A, 365, L162 

\reference Catanese, M., \& Sambruna, R.M.\ 2000, ApJ, 534, L39  

\reference Catanese, M., et al.\ 1997, ApJ, 487, L143  

\reference Chadwick, P.M., et al.\ 1999, ApJ, 513, 161 

\reference Chiaberge, M., \& Ghisellini, G.\ 1999, MNRAS, 306, 551

\reference Chiappetti, L., et al.\ 1999, ApJ, 521, 552 

\reference Costamante, L., et al.\ 2001, A\&A, 371, 512 

\reference Donato, D., Ghisellini, G., Tagliaferri, G., \& Fossati, G. 2001,
A\&A, 375, 739             

\reference Donea, A.-C., \& Protheroe, R.J.\ 2002, PASA, 19,

\reference Edelson, R.A., \& Nandra, K.\ 1999, ApJ, 514, 682

\reference Edelson, R.A., et al.\ 1995, ApJ, 438, 120

\reference Edelson, R.A., Griffiths, G., Markowitz, A., Sembay, S., Turner,
M.J.L., \& Warwick, R. 2001, ApJ, 554, 274                                     

\reference Fossati, G.\ 2001, in X-ray Astronomy 2000, Palermo,
September 2000, ASP Conf. Ser., ed.\ R.\ Giacconi, L.\ Stella, \&
S.\ Serio (San Francisco: ASP), in press

\reference Fossati, G., Maraschi, L., Celotti, A., Comastri, A., \&
Ghisellini, G. 1998, MNRAS, 299, 433         

\reference Fossati, G., et al.\ 2000a, ApJ, 541, 153 

\reference Fossati, G., et al.\ 2000b, ApJ, 541, 166 

\reference Georganopoulos, M., \& Marscher, A.P.\ 1998, ApJ, 506,
L11

\reference Ghisellini, G., Celotti, A., Fossati, G., Maraschi, L., \&
Comastri, A. 1998, MNRAS, 301, 451                

\reference Ghisellini, G., et al. 1999, A\&A, 348, 63 

\reference Giommi, P., Padovani, P., \& Perlman, E.\ 2000, MNRAS, 317, 743  

\reference Giommi, P., et al.\ 1998, A\&A, 333, L5 

\reference Giommi, P., Menna, M.T., Padovani, P. 1999a, MNRAS, 310, 465

\reference Giommi, P., et al.\ 1999b, A\&A, 351, 59 

\reference Guainazzi, M., Vacanti, G., Malizia, A., O'Flaherty, K.S.,
Palazzi, E., \& Parmar, A.N. 1999, A\&A, 342, 124       

\reference Hartman, R.C., et al.\ 2001, ApJ, 553, 683 

\reference Kataoka, J., et al.\ 1999, ApJ, 514, 138

\reference Kataoka, J., et al.\ 2000, ApJ, 528, 243 

\reference Kirk, J.G., Rieger, F.M., \& Mastichiadis, A.\ 1998,
A\&A, 333, 452


\reference Krawczynski, H., Coppi, P. S., Maccarone, T., \& Aharonian, F. A.
2000, A\&A, 353, 97         

\reference Krawczynski, H., et al.\ 2001, ApJ, 559, 187 

\reference Lamer, G., \& Wagner, S.J.\ 1998, A\&A, 331, L13 

\reference Lawson, A.J., McHardy, I.M., \& Marscher, A.P.\ 1999,
MNRAS, 306, 247 

\reference Malizia, A., et al.\ 2000, MNRAS, 312, 123    

\reference Mannheim, K.\ 1998, Science, 279, 684

\reference Maraschi, L. 2001, in Relativistic Astrophysics: 20th Texas
Symposium, AIP Conf. Proc. 586, Eds. J. C. Wheeler and H. Martel, 409
                                                                 
\reference Maraschi, L., \& Tavecchio, F.\ 2001, in X-ray
Astronomy 2000, Palermo, September 2000, ASP Conf. Ser., ed.\ R.\
Giacconi, L.\ Stella, \& S.\ Serio (San Francisco: ASP), in press
(astro-ph/0107566)

\reference Maraschi, L., et al.\ 1999, ApJ, 526, L81     

\reference Maraschi, L., Chiappetti, L., Fossati, G., Pian, E., \&
Tavecchio, F. 2000, AdSpR, 25, 713                               

\reference McHardy, I.\ 1999, in BL Lac Phenomenon, ASP Conf.\
Ser.\ 159, ed.\ L.O.\ Takalo, \& Aimo Sillanp\"a\"a,  155

\reference Padovani, P., \& Giommi, P.\ 1995, ApJ, 444, 567

\reference Perlman, E.S., Madejski, G., Stocke, J.T., \& Rector, T.A.
1999, ApJ, 523, L11
                          
\reference Petry, D., et al.\ 2000, ApJ, 536, 742        

\reference Pian, E., et al.\ 1998, ApJ, 492, L17         

\reference Rachen, J.P.\ 2000, astro-ph/0010289

\reference Ravasio, M., et al. 2002, A\&A, 383, 763

\reference Sambruna, R.M., et al. 1997, ApJ, 474, 639
 
\reference Sambruna, R.M.\ 1999, Frascati Workshop 1999, May
24--29 1999, Vulcano, Italy, MEMSAIt, (astro-ph/9912060)

\reference Sambruna, R.M., et al.\ 2000, ApJ, 538, 127   

\reference Sembay, S., et al.\ 1993, ApJ, 404, 112

\reference Tagliaferri, G., et al.\ 2000, A\&A, 354, 431 

\reference Tagliaferri, G., et al.\ 2001, A\&A, 368, 38 

\reference Takahashi, T., Madejski, G., \& Kubo, H.\ 1999, APh,
11, 177

\reference Takahashi, T., et al.\ 1996, ApJ, 470, L89    

\reference Takahashi, T., et al.\ 2000, ApJ, 542, L105   

\reference Tanihata, C., et al.\ 2000, ApJ, 543, 124 

\reference Tavecchio, F., Maraschi, L., \& Ghisellini, G.\ 1998,
ApJ, 509, 608

\reference Tavecchio, F., et al.\ 2001, ApJ, 554, 725    

\reference Treves, A., et al.\ 1999, Astronomische Nachrichten
320(4),  317 (astro-ph/9907416)

\reference Ulrich, M.-H., Maraschi, L., \& Urry, C.M.\ 1997,
ARA\&A, 35, 445

\reference Urry, C.M.\ 1999, APh, 11, 159 

\reference Urry, C.M., et al.\ 1993, ApJ, 411, 614

\reference Urry, C.M., et al.\ 1997, ApJ, 486, 799

\reference Vestrand, W.T., \& Sreekumar, P.\ 1999, APh, 11, 197

\reference Zhang, Y.H., et al.\ 1999, ApJ, 527, 719    


\end{document}